\documentclass[manuscript]{aastex}

\begin{document}
\pagenumbering{arabic}

\title{Bar Galaxies and their Environments}

\author{Sidney van den Bergh}
\affil{Dominion Astrophysical Observatory, Herzberg Institute of Astrophysics, National Research Council of Canada, 5071 West Saanich Road, Victoria, BC, V9E 2E7, Canada}
\email{sidney.vandenbergh@nrc.ca}

\begin{abstract}
                                   
The prints of the Palomar Sky Survey, luminosity classification and radial velocities were used to assign all northern Shapley-Ames galaxies to either (1) field, (2) group,
or (3) cluster environments. This information for 930 galaxies
shows no evidence for a dependence of bar frequency on galaxy
environment.  This suggests that the formation of a bar in a disk 
galaxy is mainly determined by the properties of the parent galaxy,
rather than by the characteristics of its environment.  

\end{abstract}

\keywords{galaxies:bars - galaxies:clusters}

\section{INTRODUCTION}

As Cruzen {\it et al}. (2002) have recently emphasized ``understanding the role that environment plays in the process of galaxy formation and evolution is one of the most important problems in astrophysics".  It would be particularly interesting to know if environment plays a role in the dichotomy between (1) barred and unbarred spiral disks, and (2) between pure spirals and ringed spirals.  

Why are some spiral galaxies barred, whereas others do not exhibit
bar like structures? Data compiled by Sandage \& Tammann (1981,p.91) 
show that 26\% of all disk galaxies in the Shapley-Ames catalog exhibit
bars, whereas 74\% of all disk systems \footnote{Galaxies of Hubble types S0-Sa-Sb-Sc-Sd-Sm-Im are regarded as ordinary disk systems, whereas those of types SB0-SBa-SBb-SBc-SBd-SBm-IBm are regarded as barred disk systems.} do not. It is the purpose of the present investigation to see if environmental factors might affect the fraction of all galaxies that exhibit bars. This question is also of interest in connection with the surprising finding (van den Bergh {\it et al.} 1996) that the fraction of barred galaxies appears to drop with increasing lookback time.  Detailed simulations (van den Bergh {\it et al.} 2002) show that this effect can not be entirely explained by the effects of decreasing resolution, increasing noise, and changes in the wavelength pass band at as a function of increasing redshift.  The aim of the present investigation is to see if the decrease in the frequency of barred objects with increasing redshift might be affected by environmental factors.  In particular one needs to ask whether the frequency of bars in disk galaxies might depend on environment, i.e. is the frequency of bars for galaxies of a given Hubble type the same in clusters as it is for field galaxies.

The present investigation was based on the use of the morphological classification types given in {\it A Revised Shapley Ames Catalog of Bright Galaxies} (Sandage \& Tammann (1981).  These data represent the ``golden standard'' of galaxy classification because they are based on a uniform sample of relatively nearly galaxies observed in blue light with the large reflecting telescopes at the Palomar, Mt. Wilson and Las Campanas observatories.  These galaxies were classified by expert morphologists on the system defined by Sandage (1961).  For the present investigation it is a particular advantage that Sandage \& Tammann classify galaxies either S (normal) or SB (strongly barred).  Numerical experiments show (van den Bergh {\it et al.} 2002) that the frequency distribution of such strong bars is little affected by image resolution and noise.  On the other hand van den Bergh {\it et al.} (2002) have shown that the relative frequency of weak bars [de Vaucouleurs (1959) type SAB and van den Bergh (1960abc) type S(B)] may be strongly affected by image resolution and noise.  Such weak bars can literally get lost in the noise.  The Shapley-Ames catalog classifications by Sandage \& Tammann therefore represent a robust database for the investigation of the possible dependence of bar frequency on galaxy environment.

To determine if the frequency of bars is internally determined,
or whether it depends on environment, one first has to establish
the nature of the environments of a homogeneous sample of well-
classified galaxies. The finest sample of this type is provided
by {\it A Revised Shapley-Ames Catalog of Bright Galaxies} (Sandage \&
Tammann 1981). High quality reproductions of these images have been published by
Sandage \& Bedke (1994). Other large collections of morphological classifications of galaxies, such as the {\it The Third Reference 
Catalogue} of de Vaucouleurs {\it et al.} (1991), are based on much less homogeneous observational material. Furthermore the advantage of 
a larger database in the RC3 catalog is offset by the fact that 
the RC3 galaxies are, in the mean, more distant than those in the 
Shapley-Ames catalog. Classifications in the RC3 are therefore, on
average, based on lower resolution images than are those in {\it A 
Revised Shapley-Ames Catalog} of Sandage \& Tammann.

\section{DETERMINATION OF GALAXY ENVIRONMENT}

Assigning galaxies to a particular environment is a non-trivial
problem. For instance the {\it Uppsala General Catalogue of Galaxies}
(Nilson 1973) mentions companions and group membership in an
apparently non-systematic fashion. The catalogs by Zwicky and
his collaborators (Zwicky et al. 1961-1968) give detailed lists
of galaxy clusterings. However, this information is not very useful
for a study of Shapley-Ames galaxies because these clusters are 
typically an order of magnitude more distant than the Shapley-
Ames galaxies that Sandage \& Tammann classified. It was
therefore decided to try to assign each Shapley-Ames galaxy to
a particular environment by using three criteria: (1) inspection
of the prints of the {\it Palomar Sky Survey} for the region surrounding
each galaxy, (2) by luminosity classification of nearby galaxies
(see van den Bergh 1960a), and (3) by making use of available radial 
velocity information. It should be emphasized that assigning any individual galaxy to a particular cluster is a hazardous procedure
(e.g. Puddu {\it et al.} 2001). With great care one might nevertheless
hope to achieve environmental assignments that are meaningful in a 
statistical sense. For members of the Local Group, which are so
nearby that they are spread over the entire sky, membership was
taken from van den Bergh (2000). For some nearby groups, such as 
the M81 cluster, group membership was taken from van den Bergh (1960a).
Some uncertainty is introduced in the cluster assignment process by
the fact that the internal velocity dispersion in some nearby groups
is not small compared to the mean cluster redshift. However, redshift
is a relatively reliable distance indicator for the more distant
galaxies in the Shapley-Ames catalog. Luminosity classifications
(van den Bergh 1960abc) turn out to be useful distance estimators
over the entire distance range covered by galaxies in the Shapley-
Ames catalog.

 Because of the relatively low quality of the {\it  Palomar Sky Survey} zone at $\delta$ = -30$^{\circ}$ the present investigation was restricted to the area of the sky north of $\delta$ = -27$^{\circ}$. 
Galaxies were divided into three classes on the basis of their 
environments: (1) F = Field, (2) G = group \footnote{Members of the Local Group, which are widely dispersed over the sky, were all assigned to the G class. For a number of other nearby groups and clusters, which spread over a few Palomar prints, the assignment of membership was taken from van den Bergh (1960a).} and (3) C = cluster.  The assignment of galaxies to groups and clusters was a qualitative procedure that involved both luminosity classifications using the criteria of van den Bergh (1960bc), and the distribution of group and cluster galaxies over the sky.  Companions of relatively faint distant Shapley-Ames galaxies were expected to be confined to a small fraction of the area of a single {\it Palamar Sky Survey} print, whereas the groups and clusters surrounding a bright {\it Shapley-Ames} galaxy might be spread over one or more {\it Palomar Sky Survey} prints.  

A galaxy was regarded as a member of a group if it appeared to 
contain at least three other (non-dwarf) members. A galaxy had to have 
at least six (non-dwarf) members to be assigned to the cluster category.  Furthermore clusters were assigned to S, S+E and E categories on the 
basis of their galaxian content, with S denoting a cluster in which 
most galaxies were spirals. Clusters with a mixed population were 
denoted E+S, and clusters that consisted mainly of elliptical galaxies
were denoted by E. On this system the Local Group was regarded as a 
group (G) with a mixed (E+S) population. A weaknesses of the adopted 
classification procedure is, of course, that foreground (or background)
galaxies may appear projected on a cluster without being physically 
associated with it. Since the``field'' occupies a much larger area on 
the sky than do all clusters combined, the probability of a foreground 
galaxy being projected on a rich cluster is relatively small. Such 
misassignments would add some noise to any correlations 
between environment, galaxy type and the existence of bars. An other
possible source of bias is that faint companions of distant galaxies 
near the magnitude limit of the Shapley-Ames survey may have been 
missed on the {\it Sky Survey} prints. As a result some of the most distant cluster members might have been assigned to groups, or even to the field.
   
The data on which the present investigation was based are listed 
in Table 1 (field), Table 2 (groups) and in Table 3 (clusters). The 
galaxy classifications given in these tabulation were taken from 
Sandage \& Tammann (1981), but suppress secondary information, such as 
the luminosity classifications, degree of flattening of ellipticals, 
the presence of rings, etc.  The data in Tables 1, 2, and 3 contain 
information on a total of 391 field galaxies, 145 members of groups, 
and 394 cluster members that are located in the Shapley-Ames (1932)
catalog north of $\delta$ = -27$^{\circ}$.

\section{SIGNIFICANCE OF ENVIRONMENTAL ASSIGNMENTS}

Since the  present assignment of galaxies to field, group, and 
cluster environments is based on qualitative criteria it is
important to see if these environments exhibit clear cut statistical
differences in their galactic populations. Data on this point are 
collected in Table 4. This table lists the frequency with which
different morphological galaxy types occur in differing environments. These data clearly show that early-type galaxies occur predominantly in 
groups and clusters, whereas late-type galaxies predominate in the 
field. The relative frequency of ellipticals is found to be $\sim 3$ times as great in the cluster areas as it is in the field areas. This 
reflects the well known dependence of galaxy morphology on 
environmental galaxy density (Hubble 1936, Dressler 1980). A 
Kolmogorov-Smirnov test shows that the probability that the frequency 
distribution of cluster galaxies along the sequence S0-Sa-Sb-Sc-Sd-Sm-Im 
and SB0-SBa-SBb-SBc-SBd-SBm-IBm was drawn from the same parent
population as that for field galaxies only 0.3\% . This shows beyond reasonable doubt that the adopted procedures for the assignment of galaixes to three different types different of environments has been highly successful in isolating statistically different environments.

Kraan-Korteweg \& Tammann (1979) have assigned galaxies which they suspected of having distances $<$ 10 Mpc to three environmental
classes: (1) field, (2) groups and (3) the Virgo cluster. This
differs somewhat from our own environmental assignments since some
of the richest ``groups'' of Sandage \& Tammann were designated as
``clusters'' in the present investigation. Nevertheless it is of
interest to compare the environmental assignments for the 67 objects
that are common to these two investigations. Such a comparison is
shown in Table 5. Of 10 galaxies designated as field objects in the
present investigation, Kraan-Korteweg \& Tammann assign 8 (80\%) to
the field. Of 21 galaxies assigned to groups 20 (95\%) are also
called group members by Kraan-Korteweg and Tammann. Finally of the
36 galaxies assigned to clusters in the present investigation
Kraan-Korteweg \& Tammann assign 14 (39\%) to groups and 19 (53\%)
to the Virgo cluster. These results show that the present
environmental assignments are broadly consistent with those of
Kraan-Korteweg \& Tammann.

\section{DEPENDENCE OF BAR FREQUENCY ON ENVIRONMENT}

The Revised Shapley-Ames Catalog of Sandage \& Tammann (1981)
presents an almost ideal sample for the study of the frequency 
of bars in different environments. This is so because:
(1) The Shapley-Ames galaxies are bright and therefore, on average, 
relatively nearby. As a result resolution effects on the visibility 
of bars are minimized. (2) The Shapley-Ames galaxies were classified 
on a homogeneous sample of blue sensitive plates obtained with the large
reflectors at the Palomar, Mt. Wilson and Las Campanas Observatories.
(3) These images were classified in a uniform way by two highly
competent morphologists. (4) On the morphological system of Sandage
(1961), which was employed by Sandage \& Tammann (1981), Galaxies
are classified as being either normal (S), or barred (SB). As a result
one need not worry about the possible contamination of the sample
by objects exhibiting the week bars that de Vaucouleurs (1959) assigns
to class SAB. For a detailed discussion of the effects of resolution
on the visibility of bars the reader is referred to Abraham \&
Merrifield (2000) and to van den Bergh {\it et al.} (2002).

From the data in Table 4 it is found that the frequency of bars
for galaxies located in field areas is 25 $\pm 3\%$, compared to 19  $\pm 4\%$ for galaxies in groups, and 28 $\pm$ 3\% for galaxies in clusters. These results suggest that the frequency of bar formation does not depend
significantly on galaxy environment. A caveat is, however, that the
relative frequency of various galaxy types itself depends on galaxy
environment. As a result any systematic dependence of the fraction of 
barred galaxies on Hubble type might mimic a dependence of bar frequency
on environment. Table 6 suggests that there is, in fact, little or 
no systematic variation of bar frequency along the Hubble sequence. 
As a result the systematic changes in the frequency of galaxy type 
with environment is not expected to strongly affect the dependence of
the fraction of barred galaxies on environment. This problem can be
avoided entirely by comparing the relative frequencies of barred
and normal spirals of a single Hubble stage. For galaxies of types
Sc and SBc the data in Table 4 show that the fraction of barred 
objects ranges from  17 $\pm$ 3\% in the field, through 12 $\pm$ 5\% in groups
to 25 $\pm$ 5\% in clusters. [For the entire {\it Shapley-Ames Catalog} 
Sandage \& Tammann find that 22 $\pm$ 2\% of Sc galaxies are barred.] It is tentatively concluded that presently available data do not show a 
clear-cut dependence of the fraction of barred galaxies on environment.

\section{CONCLUSIONS}

Inspection of the environments of 930 northern Shapley-Ames
galaxies on the prints of the {\it Palomar Sky Survey} has made it
possible to tentatively assign each of these objects to ``field'', ``group'' or ``cluster'' environments. No significant differences are 
found between the ratios of barred to non-barrred objects in these
different environments. These results suggest that the formation of 
bars in galaxies is mainly determined by internal factors, rather 
than by environmental effects.

I am indebted to Roberto Abraham for reawakening my interest in
the possible dependence of bar strength on galactic environment. It
is also a pleasure to thank Russell Redman for his kind help with
the formatting of the tabular material used in this investigation.  Note that the full tables 1 through 3 will be available on-line.

\begin{deluxetable}{lll}
\tablewidth{0cm}

\tablecaption{NORTHERN SHAPLEY-AMES FIELD GALAXIES.}

\tablehead {\colhead{Name} & \colhead{Type} & \colhead{Population}}
\startdata

$N 23$    &   Sb    &   ... \\
$N 24$    &   Sc    &   ... \\
$N 95$    &   Sc    &   ...  \\
$N 151$   &   SBbc  &   ...  \\
$N 157$   &   Sc    &   ...    \\
etc
\enddata
\end{deluxetable}

\begin{deluxetable}{lll}
\tablewidth{0pt}

\tablecaption{NORTHERN SHAPLEY-AMES GROUP GALAXIES.}
\tablehead {\colhead{Name} & \colhead{Type} & \colhead{Population}}
\startdata

$N 45$    &   Scd      &   S  \\
$N 128$   &   $S0$     &   E   \\
$N 147$   &   dE       &   E+S  \\
$N 185$   &   dE       &   E+S   \\
$N 205$   &   $S0/E5$  &   E+S    \\
etc
\enddata
\end{deluxetable}

\begin{deluxetable}{lll}
\tablewidth{0cm}

\tablecaption{NORTHERN SHAPLEY-AMES CLUSTER GALAXIES.}
\tablehead {\colhead{Name} & \colhead{Type} & \colhead{Population}}

\startdata

N 16    &    SB0   &   E  \\
N 227   &    E     &   E+S  \\
N 237   &    Sc    &   E+S  \\
N 245   &    Sbc   &   E+S  \\
N 357   &    SBa   &   E     \\

etc
\enddata
\end{deluxetable}

\begin{deluxetable}{lrrr}
\tablewidth{0cm}

\tablecaption{FREQUENCY OF MORPHOLOGICAL TYPES IN VARIOUS ENVIRONMENTS}

\tablehead {\colhead{Type} & \colhead{Field} & \colhead{Groups} & \colhead{Clusters}}

\startdata

E     &  $20$    &    $21$    &     $61$   \\
E/S0  &  $2$     &    $3$     &     $10$   \\
S0    &  $24$    &    $18$    &     $40.5$  \\
S0/a  &  $5$     &    $2$     &     $7$  \\
Sa    &  $27.5$  &    $7.5$   &     $24$  \\
Sab   &  $13$    &    $1.5$   &     $18$  \\
Sb    &  $46$    &    $16$    &     $21.5$  \\
Sbc   &  $27$    &    $7$     &     $21$  \\
Sc    &  $120$   &    $36.5$  &     $72$  \\
Scd   &  $2$     &    $2$     &     $6$  \\
Sd    &  $0$     &    $1$     &     $8$  \\
Sm    &  $3$     &    $0.5$   &     $3.5$  \\
Im    &  $0$     &    $1$     &     $1$  \\
SB0   &  $9$     &    $1$     &     $19$  \\
SB0/a &  $2$     &    $1$     &     $5$  \\
SBa   &  $9.5$   &    $4$     &     $9.5$  \\
SBab  &  $4$     &    $0.5$   &     $1$  \\
SBb   &  $12$    &    $7$     &     $14.5$  \\
SBbc  &  $24$    &    $1$     &     $11$  \\
SBc   &  $23$    &    $5$     &     $24.5$  \\
SBcd  &  $2$     &    $0$     &     $1$  \\
SBd   &  $0$     &    $1$     &     $1$  \\
SBm   &  $2$     &    $1$     &     $2$  \\
Other &  $14$    &    $5.5$   &     $12$  \\
\hline
Total &  $391$   &    $145$   &     $394$  \\

\enddata
\end{deluxetable}

\clearpage 

\begin{deluxetable}{lrrr}
\tablewidth{0cm}

\tablecaption{COMPARISON OF PRESENT ENVIRONMENTAL ASSIGNMENTS TO THOSE OF KRAAN-KORTEWEG \& TAMMANN (1979)}

\tablehead {\colhead{vdB type} & \colhead{F} & \colhead{G} & \colhead{C}}
\startdata

K-K\&T type    &          &           &     \\
Field          &    $8$   &   $1$     &   $3$  \\
Group          &    $2$   &   $20$    &   $14$  \\
Virgo          &    $0$   &   $0$     &   $19$  \\

\enddata
\end{deluxetable}

\begin{deluxetable}{lc}
\tablewidth{8cm}

\tablecaption{DEPENDENCE OF BAR FRACTION ON HUBBLE STAGE \tablenotemark{a}}

\tablehead {\colhead{Type} &  \colhead{Barred}} 

\startdata

S0+S0/a  & $31 \pm  4\%$  \\
Sa+Sab   & $25 \pm  4\%$  \\
Sb+Sbc   & $34 \pm  3\%$   \\
Sc       & $21 \pm  2\%$  \\
Scd-Im   & $30 \pm  7\%$ \\

\enddata

\tablenotetext{a}{Data from Tammann \& Sandage (1981)}

\end{deluxetable}

\clearpage

\section{APPENDIX}

RELATIVE FREQUENCY OF FIELD, GROUP, AND CLUSTER GALAXIES

The statistics in the present paper are affected by the fact that the rich Virgo cluster contributes significantly to the data on the northern Shapley-Ames galaxies.  Sandage \& Tammann (1981) find that 108 of the galaxies in the Shapley-Ames Catalog are located within 10$^{\circ}$ of the center of the Virgo cluster.  Data on the numbers of field, group, and cluster galaxies in the northern Shapley-Ames catalog with, and without, the contribution from the Virgo cluster are listed in Table 6.  This table show that, after excluding the Virgo region, 48 \% of all galaxies appear to be members of the field, compared to 18 \% located in groups, and 35 \% situated in clusters.

\begin{deluxetable}{ccccc}
\tablewidth{0cm}

\tablecaption{FREQUENCY OF FIELD, GROUP, AND CLUSTER MEMBERS}

\tablehead{\colhead{Environment} & \multicolumn{2}{c}{All galaxies} & \multicolumn{2}{c}{Virgo region excluded}  \\
 &  \colhead{No.} & \colhead{(\%)} & \colhead{No.} & \colhead{(\%)}}

\startdata

Field   & $391$ & $(42)$ &   $391$    &   $(48)$  \\
Group   & $145$ & $(16)$ &   $145$    &   $(18)$    \\
Cluster & $394$ & $(42)$ &   $286$    &   $(35)$   \\

\enddata
\end{deluxetable}

\end{document}